\documentclass[10pt]{ismd08}
\usepackage{graphicx}
\usepackage{cite,./mcite}

\usepackage{hyperref,url}

\newcommand{\be}{\begin{equation}}
\newcommand{\ee}{\end{equation}}
\newcommand{\bea}{\begin{eqnarray}}
\newcommand{\eea}{\end{eqnarray}}
\newcommand{\bi}{\begin{itemize}}
\newcommand{\ei}{\end{itemize}}
\newcommand{\ben}{\begin{enumerate}}
\newcommand{\een}{\end{enumerate}}
\newcommand{\la}{\left\langle}
\newcommand{\ra}{\right\rangle}
\newcommand{\lc}{\left[}
\newcommand{\rc}{\right]}
\newcommand{\lp}{\left(}
\newcommand{\rp}{\right)}

\def\frac#1#2{{{#1}\over {#2}}}
\def\gsim{\mathrel{\rlap{\lower4pt\hbox{\hskip1pt$\sim$}}
    \raise1pt\hbox{$>$}}}         
\def\lsim{\mathrel{\rlap{\lower4pt\hbox{\hskip1pt$\sim$}}
    \raise1pt\hbox{$<$}}}         

\newcommand{\draft}[1]{}

\begin{document}
\title{Update on Neural Network Parton Distributions: 
NNPDF1.1}
\author{Juan~Rojo$^{1,3}$, 
Richard~D.~Ball$^{2}$,
 Luigi~Del~Debbio$^2$, Stefano~Forte$^3$, Alberto~Guffanti$^4$, 
Jos\'e~I.~Latorre$^5$, Andrea~Piccione$^{3}$, 
 and Maria~Ubiali$^2$ }
\institute{~$^1$ LPTHE, CNRS UMR 7589, Universit\'es Paris VI-Paris VII,\\
F-75252, Paris Cedex 05, France \\
~$^2$ School of Physics and Astronomy, University of Edinburgh,\\
JCMB, KB, Mayfield Rd, Edinburgh EH9 3JZ, Scotland\\
~$^3$ Dipartimento di Fisica, Universit\`a di Milano and
INFN, Sezione di Milano,\\ Via Celoria 16, I-20133 Milano, Italy\\
~$^4$  Physikalisches Institut, Albert-Ludwigs-Universit\"at Freiburg
\\ Hermann-Herder-Stra\ss e 3, D-79104 Freiburg i. B., Germany  \\
~$^5$ Departament d'Estructura i Constituents de la Mat\`eria, 
Universitat de Barcelona,\\ Diagonal 647, E-08028 Barcelona, Spain\\
}
\maketitle
\begin{abstract}
We present recent progress within the NNPDF parton analysis framework.
After a brief review of the
results from the DIS NNPDF analysis, NNPDF1.0,
we discuss results from an updated analysis with
independent parametrizations for the strange and
anti-strange distributions, denoted by
 NNPDF1.1. We examine the phenomenological
implications of this improved analysis for the
strange PDFs.
\end{abstract}

\paragraph{Introduction}

 PDFs and  their associated uncertainties will play a crucial
 role in the full exploitation of the LHC
  physics
 potential. 
However, it is known that 
the standard approach to PDF 
determination ~\cite{Martin:2002aw,Nadolsky:2008zw}
 suffers from several drawbacks, mainly related to the lack of control
on the bias introduced in the choices of  specific PDF
parametrizations and flavour assumptions, as
well as to the difficulty in providing a
consistent statistical interpretation of PDFs uncertainties 
in the presence of incompatible data.

Motivated by this situation,
a novel method has been introduced which combines 
a MC sampling of experimental data with neural networks
as unbiased interpolators for the PDF parametrization.
This method, proposed by the NNPDF Collaboration, was first 
successfully applied to the parametrization of DIS structure functions 
\cite{f2ns,f2p} and more recently
to the determination of PDFs~\cite{DelDebbio:2007ee,Ball:2008by}. In this
contribution we present recent results within this NNPDF
analysis framework.

\paragraph{The NNPDF1.0 analysis}
NNPDF1.0 \cite{Ball:2008by} was the first DIS PDF analysis from the
NNPDF Collaboration.
The experimental dataset used in the NNPDF1.0 analysis 
consists of all relevant fixed target and
 collider deep-inelastic
scattering data: structure functions from NMC, SLAC and BCDMS,
 CC and NC reduced cross-sections from HERA, direct
$F_L(x,Q^2)$ from H1 and
 neutrino CC reduced cross sections from
CHORUS.

In NNPDF1.0, five
PDFs are parametrized with neural networks
at the initial evolution scale, which is taken to be $Q_0^2=m_c^2=2$ GeV$^2$:
$\Sigma(x,Q_0^2)$,$\quad V(x,Q_0^2)\equiv (u_v + d_v + s_v)(x,Q_0^2)$, 
$T_3(x,Q_0^2) \equiv  (u+\bar{u} -d-\bar{d})(x,Q_0^2)$,
$\Delta_S(x,Q_0^2)\equiv ( \bar{d} -\bar{u})(x,Q_0^2)$, 
and $g(x,Q_0^2)$. The  strange distributions are fixed by 
the additional assumption:
\be
\label{eq:strange}
s(x,Q_0^2)=\bar{s}(x,Q_0^2)=C_S/2\lp 
\bar{u}(x,Q_0^2)+\bar{d}(x,Q_0^2) \rp \ .
\ee
The fraction of (symmetric) strange over non-strange sea is taken to be
$C_{S}= 0.5$, as suggested by  di-muon data. While
recent analysis (see \cite{Mason:2006qa} and references therein)
suggest a somewhat smaller central value, Eq. \ref{eq:strange}
is a very crude approximation and therefore uncertainties in
$C_S$ are expected to be rather large, as the new NNPDF1.1 analysis
confirms below.

The overall normalization of  $g(x),\Delta_S(x)$ and $g(x)$ 
is fixed by imposing the momentum and valence sum rules.
The NNPDF NLO evolution  program 
employs a hybrid N-space and x-space method \cite{DelDebbio:2007ee},
whose accuracy has been checked
with
the Les Houches benchmark tables \cite{heralhc}, obtained 
from a comparison of  the {\tt HOPPET} \cite{Salam:2008qg}
and {\tt PEGASUS} \cite{pegasus} evolution programs.

\begin{figure}[ht!]
\begin{center}
\includegraphics[width=0.45\textwidth]{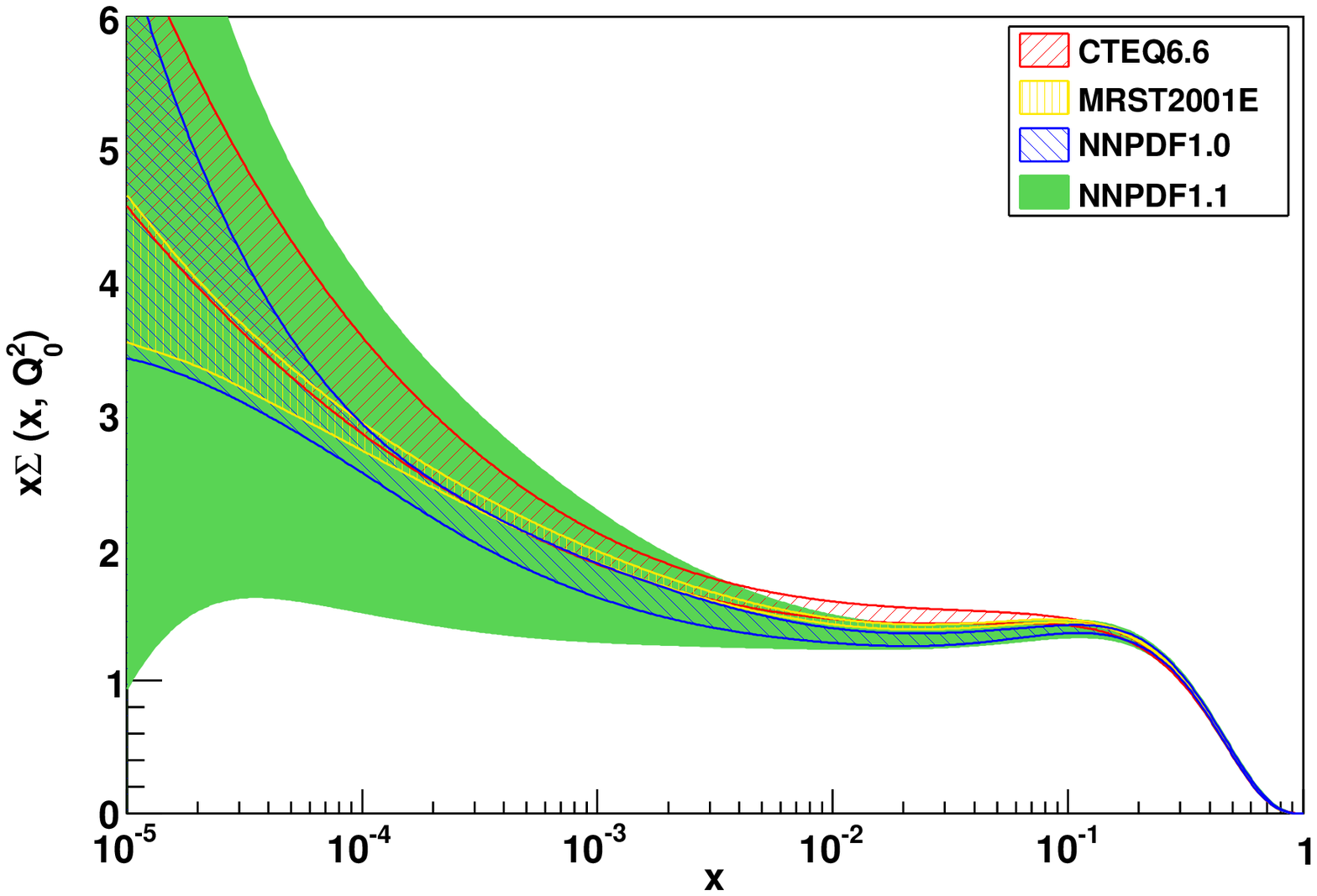}
\includegraphics[width=0.45\textwidth]{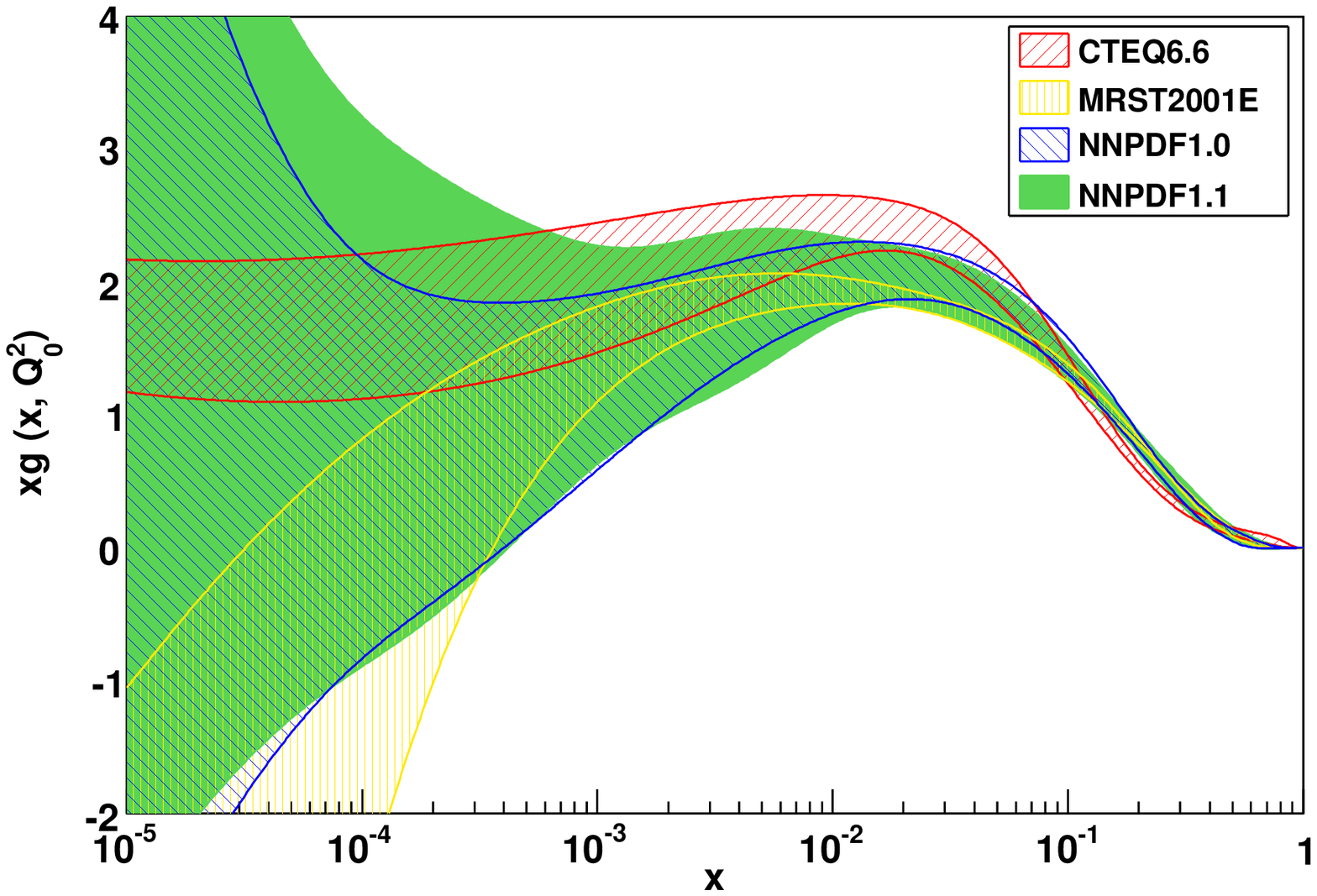}
\includegraphics[width=0.45\textwidth]{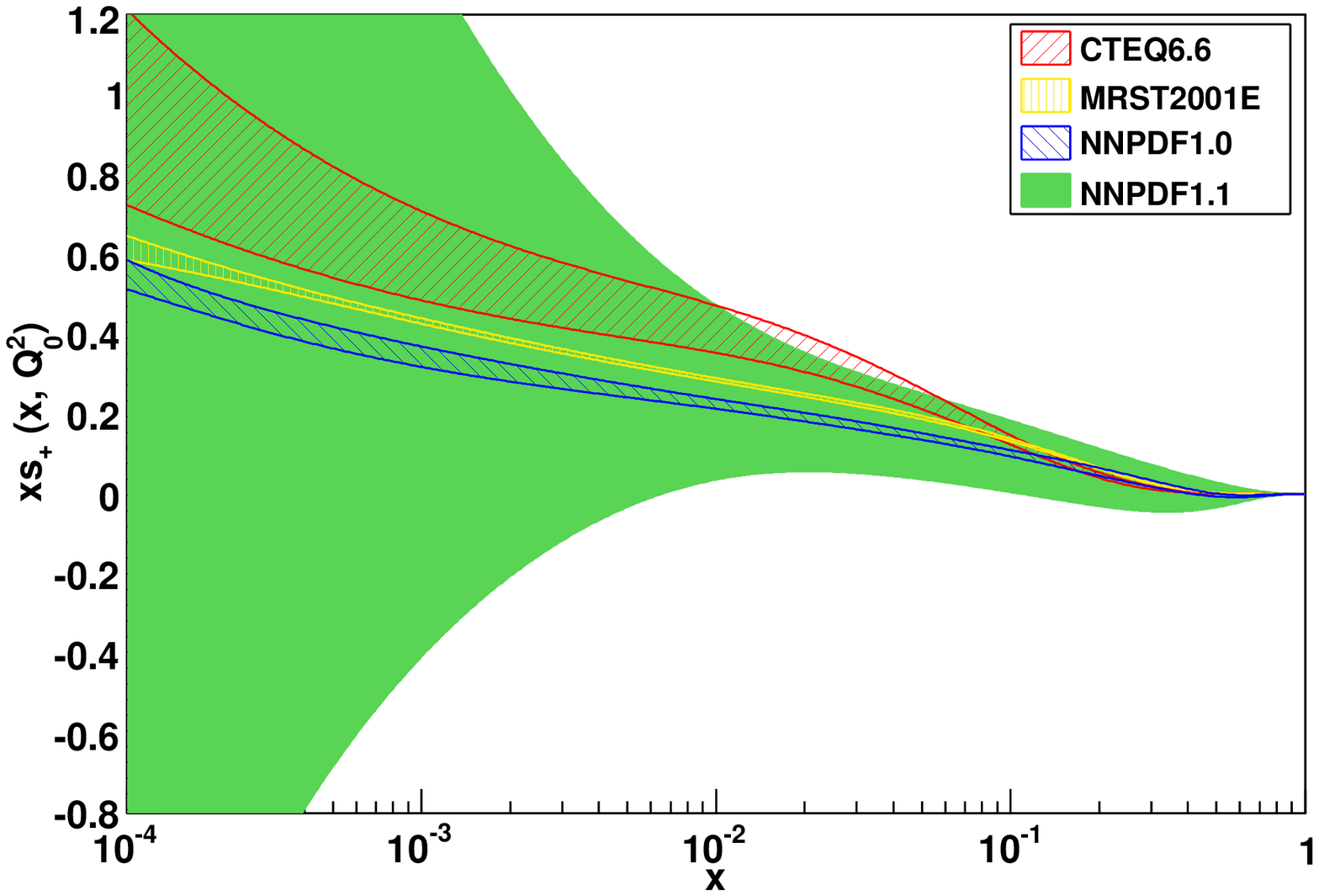}
\includegraphics[width=0.45\textwidth]{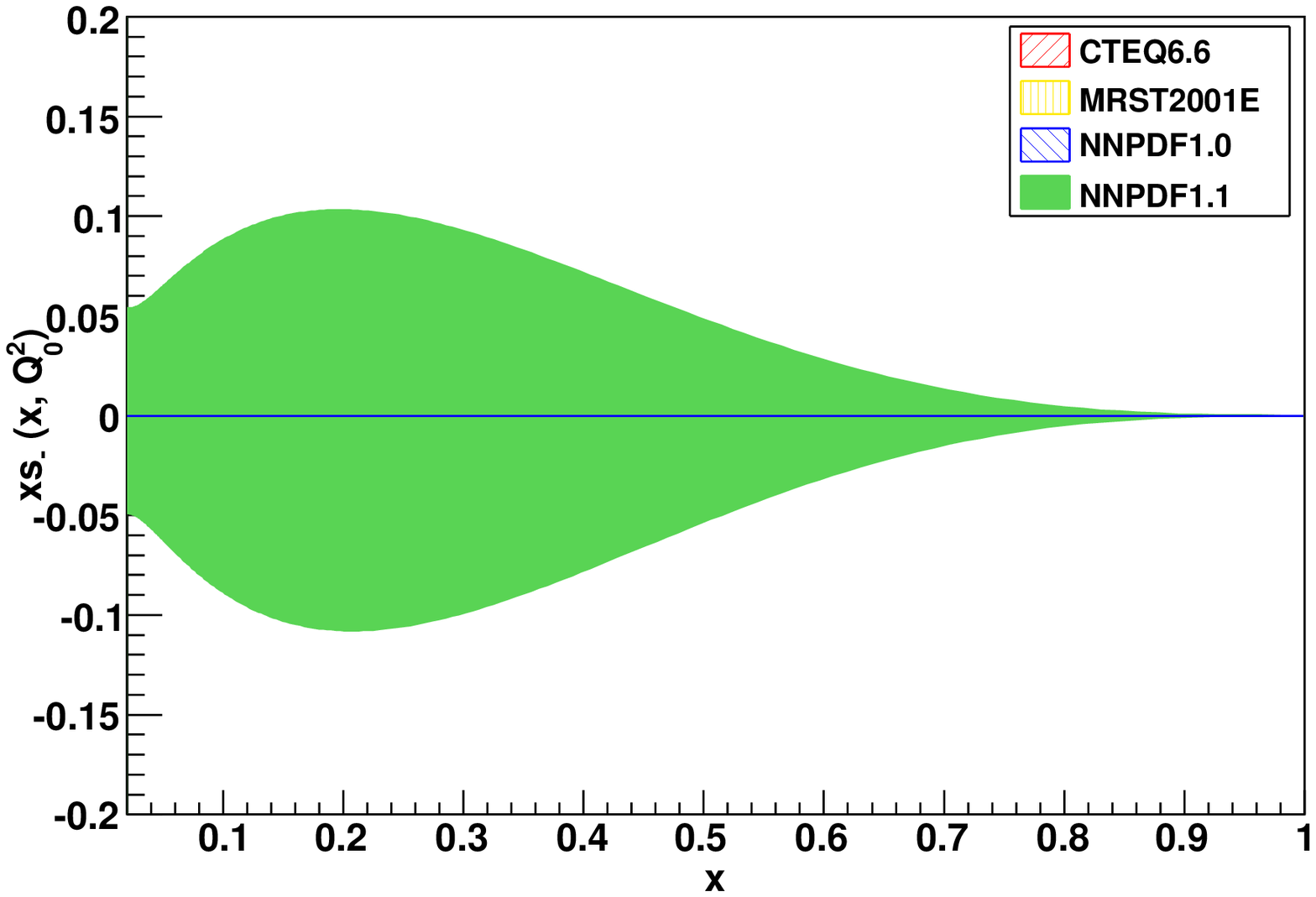}
\caption{\small The NNPDF1.1 PDFs 
compared with  other PDF sets, including NNPDF1.0.
}
\label{fig:plot-strange}
\end{center}
\end{figure}

The NNPDF1.0
gluon and singlet PDFs are shown in Fig.~\ref{fig:plot-strange},
compared with the results of other sets. 
We observe that our analysis produces results consistent 
with those obtained by  other 
collaborations~\cite{Martin:2002aw,Nadolsky:2008zw}
 while our error bands tend to get bigger in the region
 where data do not constrain PDFs behavior. Interestingly,
this happens without any error blow-up from the use of large
tolerance factors ~\cite{Martin:2002aw,Nadolsky:2008zw} in the
PDF error definition.

\paragraph{The NNPDF1.1 analysis}
NNPDF1.1 is an update of the previously
described NNPDF1.0 analysis which introduces
independent parametrizations
in the strange PDF sector and a randomization of the preprocessing.
The motivations for this update are twofold. First of all, 
the stability analysis of  \cite{Ball:2008by}, were the 
preprocessing exponents were varied their optimal values,
indicated that uncertainties might have been slightly underestimated
for some PDFs in some restricted $x-$regions, like for example the
valence PDF in the large-$x$ region. On the other hand, 
the restrictive assumptions on the strange distributions Eq. \ref{eq:strange}
should also lead to an uncertainty underestimation for some PDFs and
some observables, especially those directly sensitive to the strange
sector.

Instead
of the simplified assumptions in Eq.~\ref{eq:strange}, 
in NNPDF1.1  both $s_+(x,Q_0^2)$ and $s_-(x,Q_0^2)$
are parametrized with  independent neural networks. The 
architecture is the same as in \cite{Ball:2008by}, so that each PDF
is described by 37 free parameters.
The $s_-(x)$ distribution is forced to satisfy the strange valence sum rule
following the method of \cite{Ball:2008by}.
These strange PDFs are mostly constrained in our analysis by the CHORUS data
as well as by the HERA CC data. 

Another improvement in NNPDF1.1 with respect
to NNPDF1.0  is
a randomization of the preprocessing exponents, which
were kept fixed at their optimal values in
 \cite{Ball:2008by}. In the present analysis for each replica the PDF
preprocessing exponents are allowed to vary at random within a given
range, which is given in Table  \ref{tab:prepexps}. This range is
determined as the range in which variations of the preprocessing exponents
produce no deterioration of the fit quality, see Table 11
in  \cite{Ball:2008by}.

\begin{table}
\footnotesize
  \begin{center}
    \begin{tabular}{|c|c|c|}
      \hline PDF & $m$ & $n$ \\
      \hline
      $\Sigma(x,Q_0^2)$  & $\lc 2.7,3.3\rc$ & 
      $\lc 1.1,1.3\rc$ \\
      \hline
      $g(x,Q_0^2)$  & $\lc 3.7,4.3\rc$ & 
      $\lc 1.1,1.3\rc$ \\
      \hline
      $T_3(x,Q_0^2)$  & $\lc 2.7,3.3\rc$ & 
      $\lc 0.1,0.4\rc$ \\
      \hline
      $V_T(x,Q_0^2)$  & $\lc 2.7,3.3\rc$ & 
      $\lc 0.1,0.4\rc$ \\
      \hline
      $\Delta_S(x,Q_0^2)$  & $\lc 2.7,3.3\rc$ & 
      $\lc 0,0.01\rc$ \\
 \hline
       $s_+(x,Q_0^2)$  & $\lc 2.7,3.3\rc$ & 
      $\lc 1.1,1.3\rc$ \\
      \hline
      $s_-(x,Q_0^2)$  & $\lc 2.7,3.3\rc$ & 
      $\lc 0.1,0.4\rc$ \\
      \hline
    \end{tabular}
    \caption{\small \label{tab:prepexps} The range
of variation of the preprocessing exponents
      used in  NNPDF1.1.}
  \end{center}
\end{table}

In Fig.~\ref{fig:plot-strange} we show the results from the
NNPDF1.1 analysis for the $\Sigma(x)$, $g(x)$,$s_+(x)$
and $s_-(x)$ distributions compared to other PDF sets,
including NNPDF1.0. First of all, we observe that the central values
for both PDFs are reasonably close between NNPDF1.0 and 
NNPDF1.1, thus ensuring the validity of the flavour  assumptions
in the former case.
Second, we see that  the uncertainties
in $s_+(x)$ are large,  
so that all other PDF sets are included within the
NNPDF1.1 error band, which turns out to be
much larger than for NNPDF1.0,
since there the strange sea was fixed by Eq.~\ref{eq:strange}.
The situation for
the strange valence PDF $s_-(x)$ is similar: it
turns out to be completely unconstrained from the
present data set (see Fig.~\ref{fig:plot-strange}), with 
central value compatible with zero. 

\begin{figure}[ht!]
\begin{center}
\includegraphics[width=0.44\textwidth]{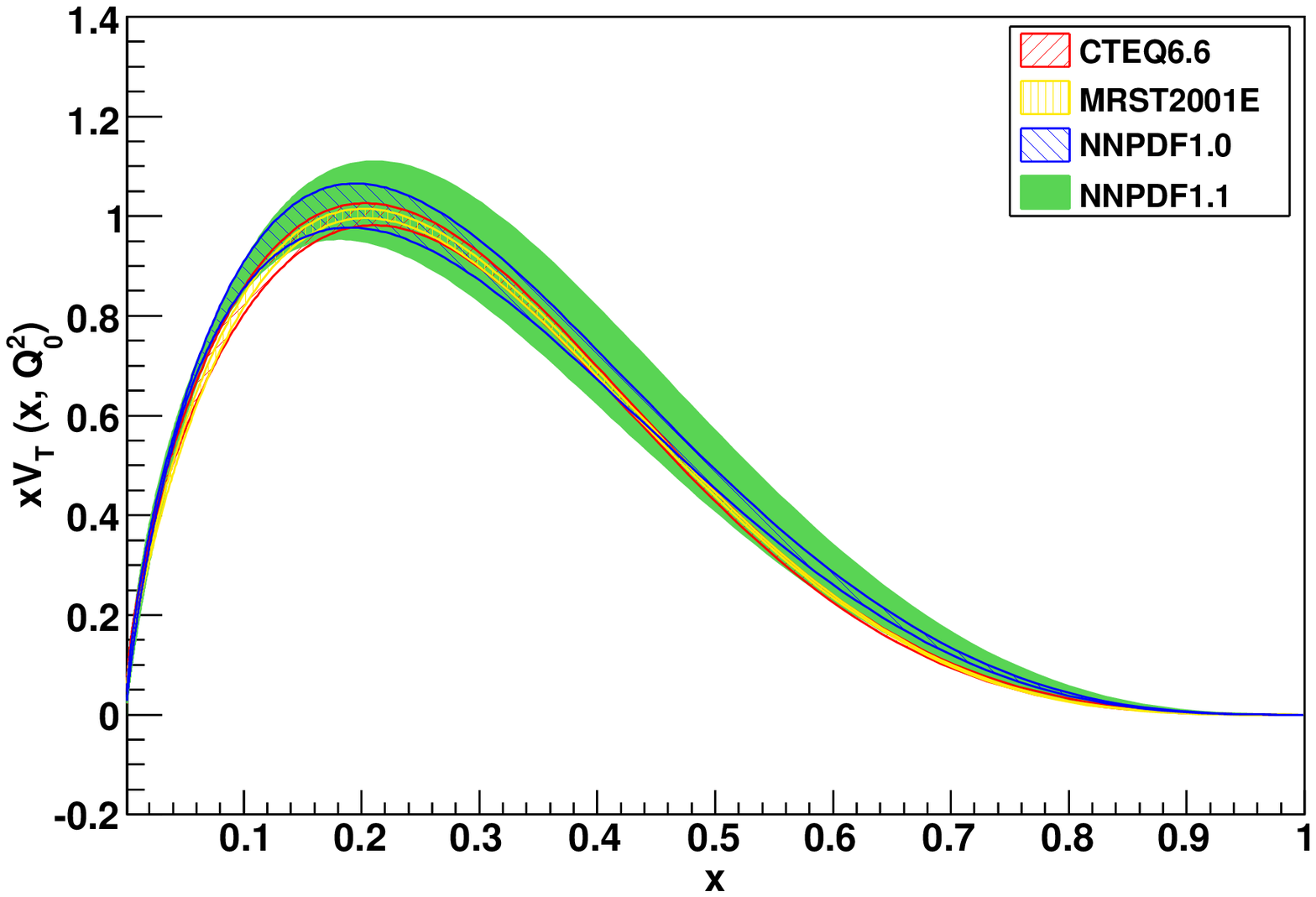}
\includegraphics[width=0.44\textwidth]{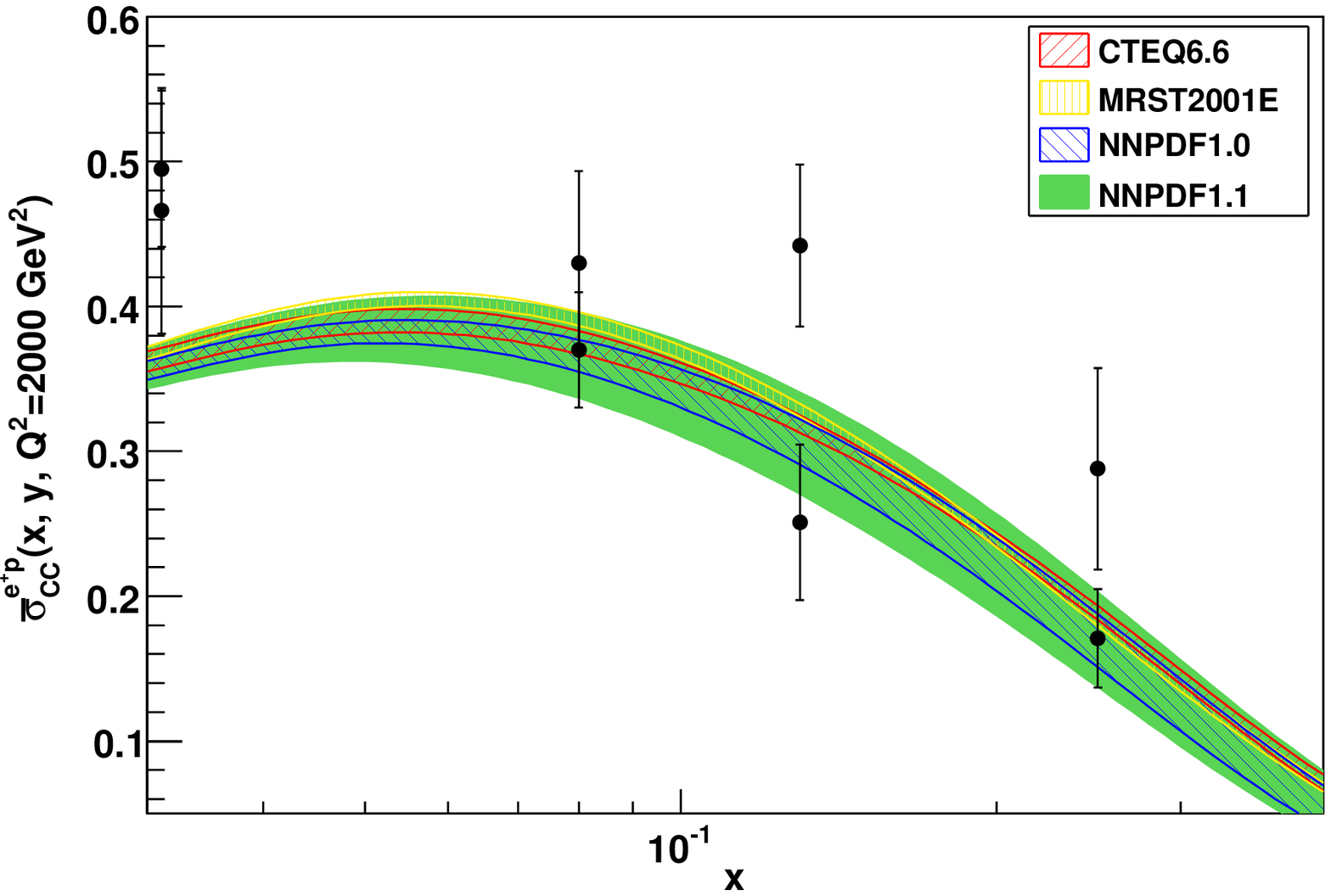}
\caption{\small The valence  PDF $V(x,Q^2_0)$ (left) and the the CC reduced cross section $\widetilde{\sigma}_{\rm CC}$ from HERA (right).
}
\label{fig:plot-xsec}
\end{center}
\end{figure}

The PDFs from the  NNPDF1.1 analysis are seen to be reasonably  
stable with respect the NNPDF1.0 ones (see $\Sigma(x)$ and $g(x)$
in Fig.~\ref{fig:plot-strange}), 
which is an important result since both  two new input PDFs
and a randomization of the preprocessing
have been incorporated in the new analysis. This stability is quantified by the
stability estimators  \cite{DelDebbio:2007ee}, shown in
Table~\ref{fig:plot-strange}. The only differences turn
out to be for
the  uncertainty in $V(x)$ 
 and on the singlet PDF $\Sigma(x)$ in the extrapolation
region. The uncertainty in 
$V(x)$, which 
 was known to be underestimated by a factor 1.5-2 in NNPDF1.0 
\cite{Ball:2008by}, now
has correspondingly increased by the expected factor, mainly due
to the absence of assumptions 
on the 
valence strange PDF $s_-(x)$, as can be seen in Fig.~\ref{fig:plot-xsec}.
A comparable increase in uncertainty is observed in
the extrapolation region of $\Sigma(x)$, which
can be attributed to the extra flexibility induced by the presence
of the independent $s_+(x)$ PDF.

\begin{table}[t!]
\begin{center}
\tiny
\begin{tabular}{|c|c|c|}
\hline 
 &  Data  & Extrapolation \\
\hline 
$\Sigma(x,Q_0^2)$ & $5~10^{-4} \le x \le 0.1$  & $10^{-5} \le x \le 10^{-4}$  \\
\hline
$\la d[q]\ra$  & 1.6  &   1.0\\
$\la d[\sigma]\ra$  & 3.5 & 2.7 \\
\hline 
\hline 
$g(x,Q_0^2)$ & $5~10^{-4} \le x \le 0.1$  & $10^{-5} \le x \le 10^{-4}$  \\
\hline
$\la d[q]\ra$  & 2.9  &  3.3 \\
$\la d[\sigma]\ra$  & 1.5 & 1.5 \\
\hline 
\hline 
$T_3(x,Q_0^2)$ & $0.05 \le x \le 0.75$  & $10^{-3} \le x \le 10^{-2}$  \\
\hline
$\la d[q]\ra$  & 1.3  &  0.9\\
$\la d[\sigma]\ra$  &  1.4 & 2.6 \\
\hline 
\hline 
$V(x,Q_0^2)$ & $0.1 \le x \le 0.6$  & $3~10^{-3} \le x \le 3~10^{-2}$  \\
\hline
$\la d[q]\ra$  & 2.2  & 2.4 \\
$\la d[\sigma]\ra$  &  5.3 & 5.3 \\
\hline 
\hline 
$\Delta_S(x,Q_0^2)$ & $0.1 \le x \le 0.6$  & $3~10^{-3} \le x \le 3~10^{-2}$  \\
\hline
$\la d[q]\ra$  & 1.0 & 1.4\\
$\la d[\sigma]\ra$  & 1.9  &1.6  \\
\hline 
\end{tabular}

\end{center}
\caption{\small Stability estimators 
which compare parton distributions from NNPDF1.0 and
NNPDF1.1. 
\label{tab:stabtab-newflav}}
\end{table}

As a consequence of the large uncertainties for $s_-(x)$,
the uncertainties in the CC DIS observables
turn out to be  larger than in NNPDF1.0,
as can
be seen in
Fig.~\ref{fig:plot-xsec}.
This result indicates that
the previously determined uncertainties in those observables
had been somewhat underestimated, as it should be obvious
by the crude assumptions concerning the
strange distributions, Eq.~\ref{eq:strange}, introduced in NNPDF1.0.

We can further study the features of the determined strange
PDFs by computing their moments.
As done for example in \cite{Lai:2007dq}, the magnitude of the strange
sea can be characterized by the following ratio of second moments:
\be
\label{eq:mom1}
K_S(Q^2)\equiv \frac{\int_0^1 dx~x~
s_+\lp x,Q^2\rp
}{\int_0^1 dx~x\lp \bar{u}\lp x,Q^2\rp+
\bar{d}\lp x,Q^2\rp\rp}= -0.1 \pm 1.7 \ ,\quad Q^2=20~ {\rm GeV}^2 \ ,
\ee
consistent within errors with the
value $C_S=0.5$ used in Eq.~\ref{eq:strange}.
On the other hand, the strange asymmetry can be characterized by
the second moment of the $s_-$ distribution, which turns out to be:
\be
\label{eq:mom2}
\la x\ra_{s_-}\equiv \int_0^1 dx~
xs_-\lp x,Q^2_0\rp = -0.001 \pm 0.04 \ ,
\ee
that is, consistent with zero within uncertainties.
This quantity has important physical implications, for example
related the determination of the Weinberg angle and 
the NuteV anomaly \cite{Davidson:2001ji}.
Both results for the strange PDF moments, Eqns.~\ref{eq:mom1} 
and \ref{eq:mom2}, further confirm the implicit
NNPDF1.0 assumption that, in the absence of further
experimental data, a PDF analysis without independent parametrizations
for $s_+$ and $s_-$ can perfectly describe all available
inclusive DIS measurements.

Our results for the moments of the
strange PDFs can be compared with 
related studies of the strange content of the nucleon
(see for example \cite{Lai:2007dq}
and references therein).
We observe that our results are compatible
with previous determinations of these moments, albeit with
large uncertainties.
These indicate that a quantitative determination of the
strange and anti-strange distributions (and the associated moments) 
requires a dedicated
study which includes experimental data directly sensitive to the
strange PDFs. The obvious example is dimuon production 
from neutrino DIS \cite{Mason:2006qa},
which is provided in a form in which it can  be consistently 
incorporated into a NLO PDF analysis.

\paragraph{Outlook}
The NNPDF1.0
DIS analysis is the first parton set within the NNPDF
framework and is
available through the LHAPDF library. The updated NNPDF1.1
analysis includes two main improvements: independent
parametrizations for the strange PDFs
and a randomization of the preprocessing. 

\paragraph{Acknowledgments}
This work has been partially supported by the
grant ANR-05-JCJC-0046-01 (France), an INFN fellowship,
as well as
by grants PRIN-2006 (Italy), and by the European network 
HEPTOOLS under contract
MRTN-CT-2006-035505.

\begin{footnotesize}
\providecommand{\etal}{et al.\xspace}
\providecommand{\href}[2]{#2}
\providecommand{\coll}{Coll.}
\catcode`\@=11
\def\@bibitem#1{%
\ifmc@bstsupport
  \mc@iftail{#1}%
    {;\newline\ignorespaces}%
    {\ifmc@first\else.\fi\orig@bibitem{#1}}
  \mc@firstfalse
\else
  \mc@iftail{#1}%
    {\ignorespaces}%
    {\orig@bibitem{#1}}%
\fi}%
\catcode`\@=12

\end{footnotesize}
\end{document}